# Mass spectrometry for semi-experimental protein structure determination and modeling

**Authors:** Gerhard Mayer[1]

1 Ruhr University Bochum, Faculty of Medicine, Medizinisches Proteom-Center, D-44801 Bochum, Germany

Gerhard Mayer

Ruhr University Bochum, Medical Faculty, Medizinisches Proteom-Center, D-44801 Bochum, Germany

Tel.: +49 234 32-21006; Fax: +49 234 32-14554; email address: gerhard.mayer@rub.de

**Abbreviations:**

| | |
|---|---|
| cryo-EM | cryo-Electron Microscopy |
| CX-MS | Cross-linking-Mass Spectrometry |
| EM | Electron Microscopy |
| HDX-MS | Hydrogen-Deuterium Exchange-Mass Spectrometry |
| IM-MS | Ion Mobility-Mass Spectrometry |
| MS | Mass Spectrometry |
| MS3D | Mass Spectrometry to derive 3D structural information |
| NMR | Nuclear Magnetic Resonance |
| PDB | Protein Data Bank |
| PPI | Protein-Protein Interaction |
| XL | Cross-Linking |
| XRC | X-Ray Crystallography |

**Summary/Abstract:**

The structure of proteins is essential for its function. The determination of protein structures is possible by experimental or predicted by computational methods, but also a combination of both approaches is possible. Here, first an overview about experimental structure determination methods with their pros and cons is given. Then we describe how mass spectrometry is useful for semi-experimental integrative protein structure determination. We review the methodology and describe software programs supporting such integrated protein structure prediction approaches, making use of distance constraints got from mass spectrometry cross-linking experiments.

## 1. Introduction

Classically one either determines the protein structure with X-ray crystallography or NMR or one tries to model the protein structure from the sequence with either homology modelling, threading or ab initio (de novo) prediction [1]. These classical experimental methods are expensive and time-consuming and have some other shortcomings as shown by the disadvantages of these experimental methods listed in Table 1. Additionally the computational methods have their limits, e.g. homology modelling requires a known template structure from a protein sequence, which has at least 30% of sequence similarity to the target sequence and the sequence-structure profile matching of threading approaches, based on statistical similarities between sequential and structural properties. Cross-linking mass spectrometry is a high-throughput experimental method, which delivers data, useful in combination for restraining the search space of such computational methods.

## 2. Overview about experimental protein structure determination methods

Since the exact prediction of protein structures in general is still an unsolved problem, protein structures until now are determined by diverse experimental methods. These experimental methods can be divided into the well-known classical methods (X-ray crystallography, NMR, cryo-EM) and the modern seminal method of coherent diffraction, which give a more or less complete set of 3D coordinates for the constituent atoms and other (spectroscopic, MS-based [2] and other structurally important experimental data generating) methods. These methods give only partial structural information and hints, which somehow correlate with the protein structure. Many of them are spectroscopic methods, which either give global information and/or information about secondary structures, or information about dynamical structural changes, e.g. due to ligand binding. These experimental data are useful for deriving distance constraints for guiding computational structure prediction methods by considerably confining the configurational search space. Table 1 gives an overview of current experimental techniques, which are useful for getting protein structure related information or for protein structure determination.

Table 1: Different kinds of experimental methods for protein structure determination or for getting protein structure related information, useful as basis for semi-experimental protein structure prediction.

| Method | | |
|---|---|---|
| *Advantages* | *Disadvantages* | *References* |
| **A) Classical protein structure determination techniques** | | |
| **Crystallization and X-ray crystallography (XRC) or neutron scattering (NS)** | | |
| • method with the highest resolution | • requires crystallized proteins and therefore a larger amount of purified protein<br>• not applicable for membrane proteins<br>• labor intensive, high cost<br>• slow<br>• access to a synchrotron is advantageous, since the resolution becomes better when using higher energetic radiation<br>• gives no information about protein dynamics due to the rigid crystal | [3,4] |
| | <ul><li>lattices</li><li>loop regions are often found in unnatural conformational states (because in fixed crystals compared with the protein in solution)</li><li>images must be calculated from diffraction pattern by Fourier synthesis</li><li>Radiation damage of crystals possible</li></ul> | |
| **Nuclear magnetic resonance (NMR)** | | |
| <ul><li>can give dynamic protein flexibility information since analysis occurs in solution</li><li>yields proton-proton distance and chemical shift constraints</li></ul> | <ul><li>only for smaller proteins (mainly < 30 kDa)</li><li>only for soluble proteins</li><li>gives only ensemble information → requires lots of computational processing</li><li>modest resolution</li><li>high cost method</li><li>high concentration of pure protein required</li></ul> | [5] |
| **Cryo-Electron microscopy (cryo-EM)** | | |
| <ul><li>gives directly images of larger structural parts like domains or α-helices</li><li>higher resolution models possible by electron density map correlation, if the atomic structure is given</li></ul> | <ul><li>medium to good resolution</li><li>radiation damage possible</li></ul> | [6-8] |
| **Holographic coherent diffraction mapping with X-ray free Electron Laser (XFEL), CDI (Coherent Diffraction Imaging)** | | |
| <ul><li>requires only a single molecule</li><li>high resolution possible</li><li>requires no protein crystal for diffraction</li></ul> | <ul><li>still difficult to handle the experimental procedure</li></ul> | [9-12] |
| **SAXS/SANS and WAXS/WANS (Small-, resp. Wide Angle X-ray / Neutron Scattering)** | | |
| <ul><li>delivers information about flexible and weakly interacting proteins</li></ul> | <ul><li>gives only low resolution information (shape reconstruction, radius of gyration)</li></ul> | [13-15] |
| **B) Spectroscopic methods** | | |
| *Short description* | | |
| **CD (Circular Dichroism)** | | |
| <ul><li>fast and relatively easy measurement</li><li>gives ratios for content of α-helices, β-sheets and random coils</li><li>gives information about the secondary structure content by measuring the absorbance of polarized light (ellipticity in dependence of the wavelength λ)</li></ul> | | [16,17] |
| **FTIR (Fourier Transform Infra-Red spectroscopy)** | | |
| <ul><li>gives mainly secondary structure information</li></ul> | | [18,19] |
| **FRET (Förster Resonance Energy Transfer)** | | |
| <ul><li>mainly for determination of molecular interactions, conformational changes and quaternary structure</li></ul> | | [20] |
| **Tryptophan fluorescence** | | |
| <ul><li>gives information about the local electrostatic interactions (e.g. hydrations) of aromatic residues like tryptophan and tyrosine and allows to monitor conformational changes caused e.g. by ligand binding</li></ul> | | [21] |
| **2D-IR (Infra-Red) spectroscopy** | | |
| <ul><li>the amide I backbone vibrations give specific signatures for different secondary structural motifs allowing to study protein dynamics on small time scales and to distinguish parallel from anti-parallel β-sheets</li></ul> | | [22-24] |
| **Deep UV Raman spectroscopy** | | |
| <ul><li>can be used for characterizing secondary structure composition</li></ul> | | [25,26] |
| **EPR (Electron Paramagnetic Resonance) and SDSL (Site-Directed Spin** | | |
| **Labeling) or PELDOR / DEER (Pulsed Electron-Electron Double Resonance)** | | |
| • allows to monitor conformational changes at the backbone level (SDSL)<br>• allows to measure distances in the nanometer range (1.5-8 nm) between paramagnetic centers like e.g. amino acid radicals (PELDOR / DEER) | | [27-29] |
| **C) Mass Spectrometric based approaches** | | |
| **Protein Cross-linking (CX-MS)** | | |
| • requires only small amount of protein<br>• can analyze protein mixtures<br>• is a High-Throughput method<br>• structure determination of proteins containing flexible and disordered regions possible<br>• gets structure information also of weakly (transient) interacting complexes<br>• can uncover novel, so far unknown protein interactions with other proteins or other biomolecules (DNA/RNA, lipids, sugars)<br>• is a relatively fast method<br>• derives distance constraints for structure refinement in homology modelling or for restraining the sample space in *ab initio* modeling<br>• dead end modifications and zero-length cross-linkers can give information about solvent accessibility<br>• allows the proteome-wide profiling to uncover PPI's [30]<br>• applicable also to membrane proteins [31] | • semi-experimental method, i.e. typically it is combined with constraint-based modelling methods<br>• requires either an enrichment step for the cross-linked peptides or the use of isotope-labelled cross-linkers<br>• identification of the cross-links has in general quadratic complexity (exception: when using isotope-labelled cross-linkers)<br>• CX-MS is most sensitive for highly abundant proteins. Therefore should be combined with pre-fractionation, e.g. size-exclusion chromatography or other enrichment methods, e.g. the use of affinity-tagged cross-linkers | [32-37] |
| **HDX-MS (Hydrogen-Deuterium Exchange-Mass Spectrometry)** | | |
| • targets hydrogens of amide groups from the protein backbone<br>• gives information about solvent accessibility and hydrogen bond status, which is dependent of secondary structure<br>• continuous labeling can give information about the different conformations of a protein<br>• pulse labeling allows to determine the influence of ligand binding on dynamic conformation changes | • requires handling with radioactive material<br>• back-exchange of deuterium limits precision of the measurements<br>• no structure information about protein core | [38-41] |
| **IM-MS (Ion Mobility-Mass Spectrometry)** | | |
| • multi-dimensional separation that allows to separate the components of protein complexes by size and shape in gas phase<br>• proteins are studied in a native-like structure state (since in gas phase)<br>• allows to determine the cross-section and radius of gyration of proteins complexes | • gives only global structural information (no detailed resolution at secondary structure or residual level) | [42-44] |
| **Native MS** | | |
| • gives quaternary structure information (stoichiometry and topology of protein complexes) | • study of membrane proteins not possible, since native MS work in aqueous solution<br>• gives only quaternary structure information (no structural information | [45-48] |

| | about the complex components) | |
|---|---|---|
| **Limited proteolysis combined with MS** | | |
| - for isolation of protein domains<br>- allows to trace conformational transitions due to ligand binding<br>- requires only low time and effort | - gives only low resolution information | [49,50] |
| **Hydroxyl radical protein footprinting (covalent labeling) combined with MS** | | |
| - gives information about solvent accessibility of the side chains of residues<br>- allows to monitor ligand-binding induced conformational changes<br>- no back exchange of labels like in HDX-MS | - relative reactivity of the residues must be taken into account | [39,51-53] |
| **Carbene footprinting** | | |
| - allows the experimental determination of the solvent accessible surface area after laser-induced photolysis of diazirine into methylene carbine :$CH_2$, which is highly reactive and has a similar size as water<br>- can be used to determine the binding sites of protein-ligand and PPI interactions | - gives only information about the surface area | [54,55] |
| **Surface Induced Dissociation (SID)** | | |
| - gives information about the connectivity, topology and the relative interface strengths of quaternary protein complexes | - can be combined with protein-protein docking approaches | [56] |
| **D) Experimental biochemical data** | | |
| **Site-directed mutagenesis experiments** | | [57] |
| - exchange of amino acid residues by site-directed nucleotide exchange in the coding gene | - experimentally laborious and therefore costly | |
| **Alanine scanning of protein surfaces** | | [58] |
| - systematic exchange of amino acid residues by alanine | - experimentally laborious and therefore costly | |
| **E) Integrative methods either combining several experimental methods or combining computational methods with experimentally derived constraints** | | |
| - combines information from several methods, especially computational and experimental methods, where the experimental data restrict the search space or guide the search | - semi-experimental method<br>- high efforts for integrating data from several sources into computational models required | [1,59-65] |

## 3. Use of data produced by mass spectrometric methods for protein structure elucidation

As shown in Table 1, the different mass spectrometric based methods can be used to get information about distance restraints [66], solvent accessibilities of residues, stoichiometry of protein complexes, the global shape and radius of gyration and also about induced conformational changes in response to binding or interaction events.

The distance constraint information got by mass spectrometric cross-linking experiments (CX-MS) can be used theoretically to determine the 3D structure of proteins and protein complexes, if a great enough variety of cross-linkers regarding both their targeted amino acid residues and their spacer lengths is available. Until now, we are not aware that a protein structure is solvable solely by CX-MS alone. But the use of cross-linking either for supplementing the information got from other experimental structure determination techniques [59,62,67,68] or in combination with computational structure prediction methods [69] to predict the protein structures in a semi-empirical way - sometimes called hybrid methods - is already routinely done [1,70]. Often a small set of distance constraints got from CX-MS can help to distinguish between two or more possible protein structures. Mass spectrometry based approaches like CX-MS, HDX-MS and IM-MS can complement classical structural biology methods and provide beside identification and quantification information also information about protein and protein complex structure and the network of interaction between the proteins. By analysis of these interaction networks hints about the cellular phenotypic states in health and disease [71] and their system behavior [72] can be derived.

The spectroscopic and mass spectrometry methods are useful to complement the information got from classical methods or can be used in cases where the classical methods are not applicable. Examples are information about disordered parts of a protein, that cannot be got by X-ray crystallography, or the structure determination of large membrane complexes, which are neither amenable by X-ray crystallography (because lack of crystals), nor by NMR (because of their size) and where cryo-EM alone can give only very low-resolution information.

Integrated structure determination methods either combine several experimental methods, or they integrate one or more of the experimental spectroscopic, biochemical and/or mass spectrometry based methods with computational structure prediction methods to give semi-empirical solved protein structures. In the following we concentrate on such semi-experimental integrative protein structure modeling methods to illustrate how the information contained in the XLMOD ontology [] can be used in CX-MS method and how it thereby can contribute to improved protein structure modeling resp. prediction.

In general the identification by a search engine is challenging, because the identification of the cross-links has in general quadratic computational complexity compared with the linear complexity of a database search in normal MS [73]. This is because one has to search for all possible binary combinations of peptides. This can either be done in an exhaustive search [74] or one explicitly encodes all the cross-linked peptides in the search database [75]. But also alternative strategies for the identification of cross-linked peptides, based on labelling of the used cross-linking reagents, were developed [76]. Beside of that also special cross-linking search engines [77,78], often used together with CID-cleavable cross-linkers [79] or isotope-labelled cross-linkers [80,81] are in use. Thereby the search can be made more efficient, because one now can easily detect the CID-fragment ion patterns [79] and / or neutral losses [34] respective the cross-linked masses by their characteristic doublets corresponding to the light resp. heavy forms of the used cross-linker [80,82]. Another method is to use cleavable cross-linkers and compare the mass shifts before and after the cleavage of the cross-linker [83]. Or one can use $^{18}$O labeling of the two oxygen atoms of the carboxyl group of the peptide C-terminus, leading to a mass shift of 8 Da for inter-peptide cross-links resp. of 4 Da for intra-peptide cross links and dead end modifications [84]. Instead of labeling the peptides one can also label the cross-linking reagents with $^{18}$O, leading to doublets separated by 4 Da for the cross-linked and the not cross-linked peptides [85]. Other cleavable cross-linkers like BuUrBu generate characteristic doublet patterns by which the cross-linked peptides can be identified [86]. PIR (Protein Interaction Reporter) is another flexible technology for the design of cross-linking reagents for distinguishing dead end, intra- and inter-peptide cross-links [87]. These PIR cross-linkers are designed with two CID labile bonds, so that a specific reporter ion, possessing an affinity or click-chemistry reactive group for purification, is released after cleavage [87]. Beside the chemically and MS cleavable cross-linkers, reviewed by Sinz [88], there are also photo-cleavable cross-linkers available, e.g. pcPIR [73]. X-links is a special search engine developed for the analysis of data sets got from such PIR cross-linking experiments [89]. Another possibility is the usage of cross-linking reagents containing MS2 labile bonds, e.g. C-S bonds, so that after MS2 cleavage the two cross-linked peptides can be unambiguously identified in the following MS3 step [90].

Another category are photoactivatable cross-linkers having an (aryl) azide, benzophenone or diazirine (resp. the isomeric diazo) reactive group [91-93]. These photoactivatable cross-linkers are mostly not selective regarding the targeted amino acid. These photoactivatable cross-linkers are very useful, because using a UV light source one can control the cross-linking reaction and one gets highly reactive intermediates resp. excited states, which react non-selectively with a great variety of amino acid residue side chains [94]. This is advantageous, since by that one gets an increased number of distance constraints, in turn providing more information for determining the protein 3D structure [95,96]. Aldehydes such as glutardialdehyde and especially formaldehyde are broadly specific, i.e. they react with many amino acid residues and also with DNA, and the cross-link yield depends on the reaction conditions and reaction times, which therefore must be carefully controlled [97]. Therefore, they allow also the study of biomolecular protein interactions in living cells. Formaldehyde can also be used for the detection of protein-DNA interactions [98] and enable one to analyze also archived formalin-fixed samples [99].

The CX-MS method provides a set of experimentally determined distance constraints, which are used for the enumeration of all protein conformations, which are in agreement with theses constraints [100]. This constraint-based protein structure modeling method can not only used for predicting the tertiary structure of a single protein, but also for the quaternary structure of protein complexes. In the protein-protein docking method normally one gets in a first search step a very large list (often tens of thousands) of conformations and then a post-docking search process is performed in order to rank these conformations got from the first step [101]. But it was shown that even after this ranking the true complex structure often cannot be found at the top of the resulting ranked list. The integration of experimentally determined distance constraints from CX-MS or other experimental methods like SAXS can help to predict the real protein complex structure by sorting out all conformations, which are not in accordance with the experimental results [59]. According to [102] in general three distance restraints are enough to determine the PPI interface. By such experimentally derived distance constraints the conformational space to be searched by protein-protein docking programs, can be drastically reduced [103], so that in combination with efficient implementations like the fast 5D-manifold Fourier Transform (FMFT) [104] the high throughput determination of PPI interfaces becomes possible.

The constraint-based approach is based on distance geometry [105] and already used in homology modeling [106], where one uses a template protein with an as high as possible sequence similarity to the target protein to model the structure of the target protein. It can also be used for fitting structures into images got from cryo-EM [107], applied to structure determination either from distances between protons, yielded by NMR measurements [108] or from distances derived from SAXS profile data [109,110]. In distance restraints got from CX-MS data, one must take into account the distances between the Cα-atoms. For instance for the cross-linking reagent BS3 with a spacer length of 11.4 Å linking two lysine residues one must add twice the distance from the ε-amino group of Lys to the Cα-atom of Lys (6.4 Å), so that the total distance becomes 11.4 Å + 2 • 6.4 Å = 24.2 Å [111]. In addition, often a tolerance of ~3-6 Å is added to account for flexibility of a dynamic backbone, so that the recommended distance to use would be 27.2 – 30.2 Å. The optimal choice of a spacer length was systematically investigated by Hofmann *et.al*, which derived the following formula for it:

$lopt[Å] = k \cdot \sqrt[3]{MW} + \sqrt[3]{SCL1 + SCL2}$, where MW is the molecular weight of the protein, SCL1 and SCL2 are the side chain lengths of the two cross-linked residues and k is a constant, which is 0.32 for Lys-Lys, 0.31 for Lys-Asp, 0.34 for Lys-Glu and Arg-Arg and 0.35 for Cys-Cys cross-links [112].

In general, with CX-MS experiments one can easily derive distance constraints for the maximum distance between two residues, whereas on must be careful in deriving minimal distance constraints, because for that one must take into account the cross-linker flexibility and the fact, that a cross-link cannot penetrate the surface of the protein [113,114].

Also one should be aware that sometimes the spacer arm cannot link two residues directly through the interior of the protein and that sometimes two residues are only not found to be linked to each other, since there are competitive suitable reactive residues at more optimal distances preferably linked [115]. Therefore, if one uses several cross-linkers with different spacer lengths, one should handle derived minimal distances with care. One possibility to handle that problem was implemented in Xwalk [113], an algorithm for validating cross-linking distances by calculating the solvent accessible surface distance (SASD) taking into account the cross-linker flexibility and that the cross-link will not penetrate the protein surface [113,114]. In addition, one can take the dynamic flexibility of the protein structure into account [116], as done by the DynamXL software. Another approach is to use specialized scoring functions, which assess beside the cross-link distances and their violations also the solvent accessibility in the scoring function [114,117].

Sometimes several distance constraints are in conflict with each other, what can e.g. be the case because of flexibility. *Ferber et.al.* developed the XL-MOD software, which can handle that in an automatic way, by using self-organizing maps, which are a special class of neural networks, to test and score different combinations of restraints [102].

The *de novo* prediction of a protein from distance constraints alone is computationally very expensive, so that it is until now realistically only for peptides and small proteins, when one has unambiguous distance constraints [100]. For cases with additional uncertainty, e.g. due to unknown side chain conformations, and for larger proteins and protein complexes one has to use approximate methods, where the side chain representations are simplified by a super atom [112]. Therefore, CX-MS is useful for integrated protein structure prediction methods, where one uses the distance constraints to refine, i.e. to filter out incorrect models from a set of predicted models [115]. The predicted models stem either from comparative (homology) modeling, to limit the sample space in *ab initio* modeling methods (which are based on molecular dynamics simulations using physical or knowledge-based, i.e. statistical force fields) [1], or are identifying the folds in threading (fold recognition) methods [118], which are based on sequence-structure profile matching.

Beside the support of tertiary structure predictions, CX-MS is an already established method for the elucidation of the quaternary structure of protein complexes. For that either restraints from CX-MS are integrated and used for improving the quality of either template-based [119] or FFT-based [103] protein-protein docking predictions or the distance restraints are used to construct a subunit interaction network with is then combined with homology modelling, as it is done by the SUMMIT algorithm [120]. CX-MS is combinable with other mass spectrometry methods for integrative protein modelling. Politis *et.al.* developed a weighted scoring function for integrating data about the overall shape (cross-section) obtained from IM-MS with stoichiometric information from native MS as well as intra-and inter-protein distance constraints from CX-MS for a restricted sampling of the conformational space [121].

## 4. Overview about tools integrating cross-linking information for protein structure prediction

Until now there are only few software programs available that support integrative structural modelling of proteins and protein complexes. Mass Spec Studio [78,122] allows the integration of data from HDX-MS [123], covalent labeling (protein footprinting) and cross-linking for the modeling of protein interactions with the protein-protein docking webserver HADDOCK (High Ambiguity driven protein-protein DOCKing) [124].

DockStar is an integrative software package for modeling of protein complexes that can integrate data from X-ray crystallography, NMR, comparative homology modeling and CX-MS [125]. It uses Integer Linear Programming (ILP) for the optimization of both knowledge-based interaction potentials as well as the satisfaction of constraints derived from CX-MS [125].

The Integrative Modeling Platform (IMP) [126] converts the information from five different experimental methods (SAXS profiles, EM2D images, EM3D density maps, the residue type content at the interface between the two proteins and distance restraints resulting from CX-MS) into a set of spatial constraints. These contraints must fulfilled simultaneously and are then optimized with a simulated annealing Monte Carlo approach followed by a refinement step [127]. This optimization based on the experimentally derived restraints drastically reduces the number of possible protein-protein docking models. IMP has several components ranging from a low level C++/Python library to user-friendly interfaces, which are integrated into the UCSF Chimera molecular visualization package [128] and allow a further integration with the homology modelling package MODELLER [129]. Another UCSF Chimera plugin useful for integrative modeling is XLinkAnalyzer [130], allowing one to integrate CX-MS data with the fitting of X-ray structures into electron density maps got from electron microscopy.

Another integrated modeling software, which can integrate distance restraints from cross-linking, is ROSETTA making use of the Xwalk algorithm [113]. It calculates the shortest path for a cross-link spacer, which leads only through solvent accessible space and not through the interior of the protein, and of XLdb, a database containing experimental cross-link data which are mapped to experimental structures from the Protein Data Bank (PDB) [131]. Kahraman *et.al.* describe 3 workflows based on

ROSETTA for data-driven comparative and for *de novo* modeling of a proteins tertiary structure, as well as for protein-protein docking.

I-TASSER (Iterative Threading ASSEmbly Refinement) [132] is a threading program for protein structure prediction, which allows to include experimentally gained distance restraints, either from cross-linking or from NMR experiments.

There are also software tools available for the visualization of cross-linking results, for instance xVis. It displays the cross links as circular, bar or network diagrams for the representation of the spatial restraints [133] (http://xvis.genzentrum.lmu.de/, accessed 02/2020), xiNET, showing interactive node-link diagrams [134] (http://crosslinkviewer.org/, accessed 02/2020) and the map viewer of XLPM [135] or VisInt-X (https://www.researchgate.net/publication/301766070_VisInt-X_Visualizing_Interactions_in_Cross-linked_proteins, accessed 02/2020). XLmap is a R package by which one can visualize contact maps with integrated of cross-link information [136].

Other interesting approaches are the specialized databases like XLinkDB 2.0 [137], integrating XL-MS data with PPI network analysis, PDB queries and homology models from the MODELLER [129] software or ProXL for visualizing, comparing, sharing, analyzing and quality control of CX-MS data [138]. MNXL [139] is a server which allows model validation based on cross-linking derived distance constraints..

Software packages supporting integrative protein structure modeling making use of cross-linking data are listed in Table 2 and a list summing up the available software tools and databases for cross-linking analysis, identification or visualization are given in Supplementary Table S1 and in the review of Yilmaz et.al. [140].

Table 2: Software packages for protein structure modeling able to integrate distance restraints derived from cross-linking data.

| Integrative modeling package | URL (accessed 02/2020) | References |
|---|---|---|
| DockStar | http://bioinfo3d.cs.tau.ac.il/DockStar/ | [125] |
| HADDOCK | http://www.bonvinlab.org/software/haddock2.2 | [124] |
| Integrative Modeling Package (IMP) | https://integrativemodeling.org | [126-128] |
| I-TASSER | http://zhanglab.ccmb.med.umich.edu/I-TASSER/ | [141] |
| Mass Spec Studio | https://www.msstudio.ca | [78,122,124] |
| MNXL | http://mnxl.ismb.lon.ac.uk/mnxl/ | [139] |
| MODELLER | https://salilab.org/modeller/ | [129] |
| ROSETTA | https://www.rosettacommons.org | [131] |
| SSEThread | https://github.com/salilab/SSEThread | [142] |
| UCSF Chimera | https://www.cgl.ucsf.edu/chimera/docs/morerefs.html | [128] |
| XLink Analyzer - UCSF Chimera plugin | http://www.beck.embl.de/XlinkAnalyzer.html | [130] |

## 5. Summary

The distance constraints got by the high-throughput CX-MS method can be used alone or in combination with constraints derived from other experimental methods to restrain, guide and accelerate the search of computational structure prediction methods. The resulting semi-experimental integrative structure prediction methodology are useful to model the structure of proteins, which cannot determined by one of the classical protein structure determination methods due to their limitations as listed in Table 1.

**Acknowledgements:**

Gerhard Mayer was funded by the BMBF grant de.NBI - German Network for Bioinformatics Infrastructure (FKZ 031 A 534A).

**Teaser:**

Cross-Linking is a high-throughput method that delivers distance constraints between two specific residues of a protein, useful as distance restraints to guide and improve computational structure modelling methods.

Supplementary Table S1: Some software packages and databases for cross-linking analysis, identification or visualization

| Name of the software package | URL (accessed 02/2020) | References |
|---|---|---|
| **Chemical drawing packages for spacer length determination** | | |
| BIOVIA Draw | http://accelrys.com/resource-center/downloads/freeware/ | --- |
| ChemBioDraw | http://www.cambridgesoft.com/Ensemble_for_Chemistry/ChemDraw/ChemDrawProfessional/ | --- |
| CORINA | http://www2.chemie.uni-erlangen.de/services/telespec/corina/ | [1] |
| **Databases for data-driven modeling** | | |
| ProXL | http://yeastrc.org/proxl_demo/viewProject.do?project_id=1 | [2] |
| XLinkDB | http://xlinkdb.gs.washington.edu | [3] |
| xComb | https://goodlett.umaryland.edu/xcomb.php | [4] |
| XDB | --- | [5] |
| **Visualization tools** | | |
| UCSF Chimera | https://www.cgl.ucsf.edu/chimera/docs/morerefs.html | [6] |
| VisInt-X | https://www.researchgate.net/publication/301766070_VisInt-X_Visualizing_Interactions_in_Cross-linked_proteins | --- |
| xiNET | https://omictools.com/xinet-tool | [7] |
| XLMap | https://omictools.com/xlmap-tool https://github.com/mammmals/XLmap | [8] |
| xVis | https://omictools.com/xvis-tool | [9] |
| XWalk | http://www.xwalk.org | [10] |
| **XL analysis and identification (search) tools** | | |
| AnchorMS | https://omictools.com/anchorms-tool | [11] |
| ASAP | --- | [12] |
| BLinks | http://brucelab.gs.washington.edu/BLinks.php | [13] |
| CLMSVault | https://gitlab.com/courcelm/clmsvault | [14] |
| CLPM | http://www.cpan.org/modules/by-module/BioX/BioX-CLPM-0.01.readme | [15] |
| Crossfinder | http://doi.org/10.1093/bioinformatics/btv083 | [16] |
| CrossSearch | --- | [17] |
| CrossWork | --- | [18] |
| Crux | http://cruxtoolkit.sourceforge.net/crux-search-for-xlinks.html | [19] |
| CTB (CoolToolBox), Virtual MSLab | ldk@science.uva.nl | [20] |
| DynamXL | http://dynamxl.chem.ox.ac.uk | [21] |
| DXMSMS | http://www.creativemolecules.com/cm_software.htm | [22] |
| ECL | http://bioinformatics.ust.hk/ecl.html | [23] |
| FindLink | --- | --- |
| FindX | http://findxlinks.blogspot.de | [24] |
| ProteinXXX / GPMAW | http://www.gpmaw.com/html/cross-linking.html | [25] |
| Hekate | http://andrewholding.com/research/hekate/ | [26] |
| Kojak | http://www.kojak-ms.org | [27] |
| ICC-CLASS (Isotopically-Coded Cleavable Cross-Linking Analysis Software) | http://www.creativemolecules.com/cm_software.htm | [28] |
| JWalk | http://jwalk.ismb.lon.ac.uk/jwalk | [29, 30] |
| MasPy | https://pypi.python.org/pypi/maspy | --- |
| MassAI | http://www.massai.dk | [31] |
| MassMatrix | http://www.massmatrix.net | [32] |
| MaxQuant | http://www.coxdocs.org/doku.php?id=maxquant:start | [33] |
| MeroX | http://www.stavrox.com | [34] |
| MS2Assign | --- | [35] |
| MS2Links | --- | --- |
| MS2Pro | --- | [36] |
| MS3D | --- | --- |

| Tool | URL | Ref |
|---|---|---|
| MS-Bridge (UCSF ProteinProspector) | http://prospector.ucsf.edu/prospector/cgi-bin/msform.cgi?form=msbridgestandard | [37] |
| MXDB | https://omictools.com/mxdb-tool | [38] |
| MSX-3D | --- | [39] |
| OpenPepXL | http://www.openms.de/comp/openpepxl/ | --- |
| OpenProXL | --- | --- |
| PepSearch | https://omictools.com/pepsearch-tool | --- |
| PeptideMap (PROWL) | http://prowl.rockefeller.edu/prowl/peptidemap.html | [40] |
| pLink | https://omictools.com/plink-2-tool | [41, 42] |
| pLink-SS (disulfide bonds) | http://pfind.ict.ac.cn/software/pLink/2014/pLink-SS.html | [43] |
| Popitam | --- | [44, 45] |
| Pro-Crosslink | https://els.comotion.uw.edu/licenses/3 | [46] |
| pXtract | http://pfind.ict.ac.cn/software/pXtract/index.html | --- |
| RNPxl | http://open-ms.sourceforge.net/publications/rnpxl/ | [47] |
| SearchXLinks | | [48] |
| SIM-XL (Spectrum Identification Machine-XL) | http://patternlabforproteomics.org/sim-xl/ | [49] |
| SLinkS | https://xlinkx2beta.hecklab.com | [50] |
| StavroX | http://www.stavrox.com | [51] |
| SQID-XLink | http://chemistry.osu.edu/~wysocki.11/bioinformatics.htm | [52] |
| XFDR | ftp://ftp.mi.fu-berlin.de/pub/OpenMS/develop-documentation/html/UTILS_XFDR.html | --- |
| xiFDR | https://github.com/lutzfischer/xiFDR | [53] |
| XI | http://rappsilberlab.org/rappsilber-laboratory-home-page/tools/ | --- |
| XiQ | http://rappsilberlab.org/tools/ | [54] |
| xilmass | https://github.com/compomics/xilmass | [55] |
| XL-MOD | https://omictools.com/xl-mod-tool | [56] |
| X-Link | --- | [57] |
| X-Links | --- | [58] |
| X!Link | yjlee@iastate.edu | [12, 59] |
| XLink | --- | [60] |
| XLink Analyzer | http://www.beck.embl.de/XlinkAnalyzer.html | [61] |
| XLink-Identifier | http://www.du-lab.org | [24] |
| XLinkProphet | | [62] |
| XlinkX | https://xlinkx.hecklab.com | [63] |
| XL-MOD | http://aria.pasteur.fr/supplementary-data/x-links | [64] |
| XLPM (X-Linked Peptide Mapping Algorithm) | http://binf-app.host.ualr.edu/~mihir/cgi-bin/xlpm.cgi | [65] |
| xQuest / xBobcat | http://prottools.ethz.ch/orinner/public/htdocs/xquest/ | [66] |
| xQuest / xProphet | http://proteomics.ethz.ch/cgi-bin/xquest2_cgi/index.cgi | [67] |
| xTract | https://omictools.com/xtract-tool | [68] |
| XXXLink | --- | [31] |
| **Structure model validation** | | |
| DisVis | http://milou.science.uu.nl/services/DISVIS | [69] |
| MNXL | http://mnxl.ismb.lon.ac.uk | [30] |
| PowerFit | http://milou.science.uu.nl/services/POWERFIT | [69] |